\documentclass[reprint,aps,prl,superscriptaddress,floatfix]{revtex4-1}
\pdfoutput=1
\usepackage[utf8]{inputenc}
\usepackage[T1]{fontenc}

\usepackage{microtype}

\usepackage[single]{acro}
\usepackage{bbold}
\usepackage{braket}
\usepackage{mathtools}
\usepackage{amsfonts}
\usepackage{siunitx}
\DeclareSIUnit{\sqrthz}{\ensuremath{\sqrt{\text{\hertz}}}}
\DeclareSIUnit{\gamoveromeg}{\ensuremath{\left(\displaystyle\frac{\gamma}{\Omega}\right)}}

\newcolumntype{C}[1]{>{\centering\let\newline\\\arraybackslash\hspace{0pt}}m{#1}}

\usepackage[colorlinks=true,citecolor=red,linkcolor=brown,urlcolor=blue]{hyperref}

\DeclareAcronym{EM}{
	short = EM,
	long = electromagnetic,
}
\DeclareAcronym{GW}{
	short = GW,
	long = gravitational wave,
}
\DeclareAcronym{CFI}{
	short = CFI,
	long = classical Fisher information,
}
\DeclareAcronym{CRB}{
	short = CRB,
	long = Cramér-Rao bound,
}
\DeclareAcronym{QCRB}{
	short = QCRB,
	long = quantum Cramér-Rao bound,
}
\DeclareAcronym{QFI}{
	short = QFI,
	long = quantum Fisher information,
}
\DeclareAcronym{POVM}{
	short = POVM,
	long = positive operator valued measurement,
}
\DeclareAcronym{SNL}{
	short = SNL,
	long = shot-noise limit,
}
\DeclareAcronym{SQL}{
	short = SQL,
	long = standard quantum limit,
}

\newcommand\estimator[1]{\tilde{#1}}
\newcommand\expect[1]{\langle #1 \rangle}

\newcommand\expectation[1]{\mathbb{E}[#1]}
\newcommand\identity{\mathbb{1}}

\newcommand\op[1]{\hat{#1}}
\newcommand\opvec[1]{\mathbf{\hat{#1}}}
\newcommand\trace[1]{\mathrm{Tr}\left[ #1 \right]}

\DeclareMathOperator\diag{diag}

\begin{document}

\title{Fundamental Quantum Limits of Multicarrier Optomechanical Sensors}

\author{Dominic Branford}
\affiliation{Department of Physics, University of Warwick, Coventry CV4 7AL, United Kingdom}

\author{Haixing Miao}
\affiliation{School of Physics and Astronomy, Institute of Gravitational Wave Astronomy, University of Birmingham, Birmingham B15 2TT, United Kingdom}

\author{Animesh Datta}
\affiliation{Department of Physics, University of Warwick, Coventry CV4 7AL, United Kingdom}

\date{\today}

\begin{abstract}
Optomechanical sensors involving multiple optical carriers can experience mechanically mediated interactions causing multi-mode correlations across the optical fields. 
One instance is laser-interferometric gravitational wave detectors which introduce multiple carrier frequencies for classical sensing and control purposes.
An outstanding question is whether such multi-carrier optomechanical sensors outperform their single-carrier counterpart in terms of quantum-limited sensitivity.
We show that the best precision is achieved by a single-carrier instance of the sensor.
For the current LIGO detection system this precision is already reachable.
\end{abstract}

\maketitle

\footnotetext[1]{See supplementary material for further information, which includes Refs.~[51--54]}

\textit{Introduction.---}The use of quantum-mechanical systems and non-classical properties for high-precision estimation tasks has attracted interest in a number of sensing schemes, including in laser-interferometric \ac{GW} detectors~\cite{caves_quantum-mechanical_1981,kimble_conversion_2001,the_ligo_scientific_collaboration_gravitational_2011,demkowicz-dobrzanski_fundamental_2013,miao_towards_2017} and related problems~\cite{tsang_fundamental_2011,lang_optimal_2013}, magnetometry~\cite{sheng_subfemtotesla_2013,baumgratz_quantum_2016}, and atomic clocks~\cite{bollinger_optimal_1996,macieszczak_bayesian_2014}.
Direct detection of \acp{GW} was one of the earliest problems to demand such analysis~\cite{braginsky_quantum_1996}, suggesting use of non-classical light---squeezed vacuum states---to improve precision~\cite{caves_quantum-mechanical_1981,schnabel_squeezed_2017}.

Sensing mechanical displacements optically, such as in laser-interferometric GW detectors~\cite{braginsky_quantum_1992,danilishin_quantum_2012}, relies on interactions between optical and mechanical degrees of freedom is the domain of optomechanical~\cite{chen_macroscopic_2013,aspelmeyer_cavity_2014} sensors.
Light incident on a mechanical oscillator causes the mechanical oscillator to act as an active element which produces squeezing of optical modes~\cite{braginsky_ponderomotive_1967,kimble_conversion_2001}---the so-called ponderomotive squeezing.
Such squeezing acts as a noise source constraining the current generation of laser-interferometric \ac{GW} detectors~\cite{braginsky_quantum_1992,kimble_conversion_2001} due to anti-squeezing of the quadrature in which the signal is encoded which manifests as a measurement backaction, with techniques to avoid such backaction drawing significant interest~\cite{tsang_coherent_2010,ockeloen-korppi_quantum_2016,moller_quantum_2017}.
The same effect has been demonstrated as a squeezed light source~\cite{brooks_non-classical_2012,safavi-naeini_squeezed_2013,purdy_strong_2013}, which can potentially improve sensors' precision~\cite{caves_quantum-mechanical_1981,kimble_conversion_2001,schnabel_squeezed_2017}.

The extension to multi-mode optomechanical systems has proven fruitful in both the many mechanical~\cite{nielsen_multimode_2017} and optical~\cite{lee_multimode_2015,slatyer_synthesis_2016} mode scenarios, as well as for optical frequency conversion~\cite{hill_coherent_2012,andrews_bidirectional_2014}.
This includes sensors such as laser-interferometric \ac{GW} detectors, particularly those encompassing modifications which utilise multiple laser frequencies: so-called multi-carrier interferometers.
Originally implemented in Advanced LIGO for classical sensing and control purposes~\cite{izumi_multicolor_2012,staley_achieving_2014},
a second carrier can improve the low-frequency sensitivity by partially cancelling the strong backaction of the main carrier~\cite{rehbein_local_2007,miao_quantum_2014}.
Multiple carriers can provide a means to enhance the sensitivity and surpass the \ac{SQL}~\cite{braginsky_quantum_1992} by using the optical spring effect, while not suffering from the instabilities associated with the single-carrier case and allowing for some shaping of the sensitivity curves~\cite{rehbein_double_2008,korobko_paired_2015}.
The value of multiple carriers in improving the sensors' fundamental quantum limit, which is more stringent than the \ac{SQL}, remains open.

In this Letter we provide the fundamental quantum limits on the precision of multi-carrier optomechanical sensors, including laser-interferometric \ac{GW} detectors, using quantum metrology techniques.
These limits are imposed by the classical and quantum Fisher information---via the \ac{CRB} on precision of an estimator---from quantum estimation theory~\cite{holevo_probabilistic_2011,braunstein_statistical_1994,hayashi_asymptotic_2005,paris_quantum_2009,toth_quantum_2014,demkowicz-dobrzanski_quantum_2015}.
Our multimode analysis includes optical loss at the output and squeezed light injection; as well as the optomechanical interaction---the ponderomotive squeezing effect.

Multi-mode quantum states have been studied in quantum metrology~\cite{pinel_ultimate_2012,humphreys_quantum_2013,friis_heisenberg_2015,ciampini_quantum-enhanced_2016,gagatsos_gaussian_2016}. 
By including a noise source which itself introduces multi-mode correlations, ponderomotive squeezing, for the first time we show that for a large class of optomechanical sensors multiple carriers are no better than single carriers.
Hitherto neglected in estimation-theoretic quantum metrology studies of \ac{GW} detectors~\cite{lang_optimal_2013,demkowicz-dobrzanski_fundamental_2013}
ponderomotive squeezing dominates the low-frequency quantum noise of \ac{GW} detectors~\cite{kimble_conversion_2001} as well as smaller optomechanical systems~\cite{purdy_observation_2013,teufel_overwhelming_2016,cripe_observation_2018}.
We bridge this gap, providing analytical expressions for the fundamental quantum limits of multi-mode optomechanical sensors featuring ponderomotive squeezing.
This should guide the development of novel optomechanical sensors and the improvement of existing ones.
Our large complement of results can be navigated using Table~\ref{tab:results}.

\begin{table*}
	\centering
	\begin{ruledtabular}
	\begin{tabular}{ C{4.0cm} C{4.25cm} C{4.25cm} C{4.25cm} }
		Input \& output 					& Fundamental limit 					& Freq.\ dependent homodyne 						& Signal quadrature homodyne \\ \hline
		Squeezed \& lossy 				& Eq.~\eqref{eq:qcrb_squeezeloss} 		& Eq.~\eqref{eq:crb_squeezeloss} 					& Eq.~\eqref{eq:signalhom_crb} \\
		Identically squeezed \& lossy 	& Eq.~\eqref{eq:qcrb_equalsqueezing} 	& Eq.~\eqref{eq:qcrb_equalsqueezing}\footnotemark[1]	 & Eq.~\eqref{eq:signalhom_equalsqueezing} \\
		Unsqueezed \& lossy 			& Eq.~\eqref{eq:qcrb_unsqueezed} 		& Eq.~\eqref{eq:qcrb_unsqueezed}\footnotemark[1]		 & Eq.~\eqref{eq:signalhom_unsqueezed} \\
		Squeezed \& lossless 			& Eq.~\eqref{eq:qcrb_losslesssqueeze} 	& Eq.~\eqref{eq:qcrb_losslesssqueeze}\footnotemark[2]  		& Supplementary material~\cite{note1} \\
	\end{tabular}
	\end{ruledtabular}
	\footnotetext[1]{Attainable through the homodyne angle given by Eq.~\eqref{eq:hom_angle_equal}, otherwise for general homodyne angles these are found as limits of Eq.~\eqref{eq:crb_squeezeloss} or in Sec.~VI of the supplementary material~\cite{note1}.}
	\footnotetext[2]{Attainable through the homodyne angle given by Eq.~\eqref{eq:hom_angle_lossless}, otherwise for general homodyne angles these are given in Sec.~VI of the supplementary material~\cite{note1}.}
	\caption{
		Expressions for precision of various interferometer limits.
		The unsqueezed and lossless case can be most readily recognised from the lossy and unsqueezed case with limit \( \eta = 1 \).
	We provide some discussion of these results in the context of LIGO detector in Sec.~VIII of the Supplementary material~\cite{note1}.
	}
	\label{tab:results}
\end{table*}

\nocite{miao_quantum_2012,the_ligo_scientific_collaboration_advanced_2015,thorne_modern_2017,kay_fundamentals_1998}

\begin{figure}[htb]
	\centering
	\includegraphics{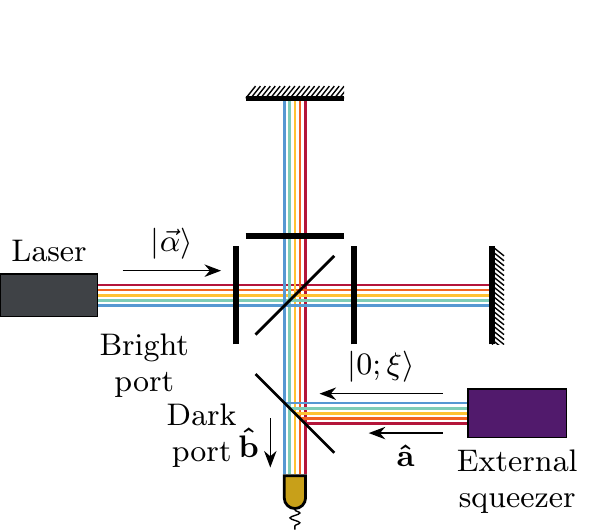}
	\caption{
	An instance of an optomechanical sensor---a laser-interferometric gravitational wave detector with multiple different frequency carrier-modes.
	Circulating light couples to the mechanical motion of the mirrors inside the interferometer arms.
	\( \opvec{a} \) and \( \opvec{b} \) describe the field of the carrier-mode sidebands, entering and exiting the interferometer respectively at the dark port.
	}
	\label{fig:interferometer}
\end{figure}

\textit{Framework.---}%
We describe the optical part of our optomechanical sensor with a linear input-output relation
\begin{equation}
	\opvec{b}(\Omega) = \mathcal{M}(\Omega) \opvec{a}(\Omega) + h(\Omega) \vec{\mathcal{V}}(\Omega),
	\label{eq:inputoutput_multimode}
\end{equation}
where \( \mathcal{M} (\Omega) \) is a complex matrix which determines a Bogoliubov transformation between the incoming and outgoing fields, and \( h(\Omega) \vec{\mathcal{V}}(\Omega) \) is a displacement vector which encodes the parameter \( h(\Omega) \).
Such input-output relations are typically expressed in terms of the two-photon formalism~\cite{caves_new_1985,schumaker_new_1985}, using the two operators
\(
	\op{a}_1^{(\omega)} = ( \op{a}_{ \omega + \Omega } + \op{a}_{ \omega - \Omega }^{\dagger} ) / \sqrt{2},
\)
and
\(
	\op{a}_2^{(\omega)} = -i ( \op{a}_{ \omega + \Omega } - \op{a}_{ \omega - \Omega }^{\dagger} ) / \sqrt{2}.
\)
We introduce \( d \) pairs of such operators \( \{ \op{a}_1^{(\omega_1)} , \op{a}_2^{(\omega_1)} , \cdots , \op{a}_2^{(\omega_d)} \} \) to describe the \ac{EM} fields in an interferometer driven by light of multiple carrier frequencies~\( \{ \omega_1, \omega_2, \cdots, \omega_d \} \).
From these creation/annihilation operators, we can form hermitian position (\(\op{x}_{1,2}^{(\omega)}\)) and momentum (\(\op{p}_{1,2}^{(\omega)}\)) operators, spanning the same phase space and obeying suitable commutation relations~\cite{note1}.

Suppressing the \( \Omega \) argument for brevity; we focus on the case where we wish to estimate the size of the displacement \( h \), with \( \mathcal{M} \) and \( \mathcal{V} \) consisting of the \( 2{\times}2 \) and \( 2{\times}1 \) blocks~\cite{rehbein_double_2008,miao_quantum_2014}, see also Sec.~II of the supplementary material~\cite{note1}
\begin{equation}
\begin{gathered}
	\mathcal{M}_{jk} =
	e^{i(\beta_j+\beta_k)}
	\begin{pmatrix}
		\delta_{jk} & 0 \\
		- \chi\sqrt{ \kappa_j \kappa_k } & \delta_{jk}
	\end{pmatrix},
	\\
	\mathcal{V}_{j} =
	\frac{ e^{i\beta_j} }{ h_{\text{SQL}} }
	\begin{pmatrix}
		0 \\
		\chi\sqrt{ 2 \kappa_j }
	\end{pmatrix},
\end{gathered}
\label{eq:multiMV_inout}
\end{equation}
where \( \delta_{ij} \) is the Kronecker delta, \( \beta_i \) are phases, \( \kappa_i \geq 0 \).
 \( \chi \in \{ -1,1 \} \) is the sign of the mechanical response and can be taken to be positive, since one with a negative response is identical to one with a positive \( \chi \) with a fixed phase shift preceding and succeeding it, which can be captured by rotating input squeezing and output homodyne angles respectively.
The attainable precisions are thus directly related; see Sec.~II of the supplementary material~\cite{note1}.
The presence of the \( \sqrt{\kappa_j \kappa_k} \) term on the off-diagonals produces a multi-mode squeezing across all the optical modes,
which is ponderomotive in origin.
The ponderomotive squeezing introduced with a single optical mode---with frequency \( \omega \)---is itself multi-mode with correlations between the \( \omega+\Omega \) and \( \omega-\Omega \).
When multiple optical fields are used they each affect the mechanical motion and in turn the mechanical motion causes squeezing of each optical mode leading to entanglement between \( \omega_j+\Omega \) and \( \omega_k+\Omega \) optical modes.

In the case of a multi-carrier laser-interferometric \ac{GW} detector as in Fig.~\ref{fig:interferometer} in the tuned configuration, \( \kappa_i \) is the normalised intensity of the \( i \)-th carrier
\begin{equation}
\begin{aligned}
	\kappa_i &= 
	\frac{ 16 I_i \omega_i \gamma_i }{ m c L \Omega^2 \left( \gamma_i^2 + \Omega^2 \right) },&
	h_{\text{SQL}} &= 
	\sqrt{ \frac{ 8 \hbar }{ m \Omega^2 L^2 } },
\end{aligned}
\label{eq:kappa_i}
\end{equation}
where \( I_i \) is the arm cavity power of the \( i \)th mode, \( \omega_i \) the frequency of the \( i \)th mode, \( \gamma_i \) the arm cavity half-bandwidth of the \( i \)th mode, \( m \) the test mass, and \( L \) the interferometer arm length~\cite{danilishin_quantum_2012}.
The signal-recycling mirror~\cite{buonanno_quantum_2001,buonanno_signal_2002,buonanno_scaling_2003,mcclelland_lsc_2017} introduces more involved input-output relations but at low-frequencies where radiation-pressure dominates the quantum noise they can be approximated with the same form of Eq.~\eqref{eq:multiMV_inout}~\cite{corbitt_squeezed-state_2006}.
Interferometer modifications such as the quantum speed meter~\cite{purdue_practical_2002,danilishin_quantum_2012,mcclelland_lsc_2017} also have the same form of input-output relations as Eq.~\eqref{eq:multiMV_inout} and our results can be applied directly with appropriate definition of \( \kappa_i \).

As Eq.~\eqref{eq:inputoutput_multimode} is a linear mapping between creation operators, the optical fields through the sensor evolve under a Gaussian unitary~\cite{adesso_continuous_2014}.
Common input states such as (squeezed) vacuum are themselves Gaussian~\cite{caves_quantum-mechanical_1981,kimble_conversion_2001,the_ligo_scientific_collaboration_gravitational_2011}, therefore the output state can be taken as Gaussian for relevant cases.
From the evolution of the quadrature operators
\begin{equation}
\begin{aligned}
	\opvec{x}' &=
	\frac{ \mathcal{M} \opvec{a} + \mathcal{M}^* \opvec{a}^{\dagger} + h \vec{\mathcal{V}} + h^* \vec{\mathcal{V}}^* }{ \sqrt{2} },
	\\
	\opvec{p}' &=
	\frac{ \mathcal{M} \opvec{a} - \mathcal{M}^* \opvec{a}^{\dagger} + h \vec{\mathcal{V}} - h^* \vec{\mathcal{V}}^* }{ i\sqrt{2} },
\end{aligned}
\end{equation}
we can extract the displacement and symplectic operators
\begin{equation}
\begin{aligned}
	\vec{d}_{\mathcal{V}} =
	\sqrt{2}
	\begin{pmatrix}
		\Re [ h \vec{\mathcal{V}} ] \\
		\Im [ h \vec{\mathcal{V}} ]
	\end{pmatrix},
	&&
	\mathcal{S}_\mathcal{M} =
	\begin{pmatrix}
		\Re \mathcal{M} & -\Im \mathcal{M} \\
		\Im \mathcal{M} & \Re \mathcal{M}
	\end{pmatrix},
\end{aligned}
\label{eq:displacement_symplectic}
\end{equation}
where \( \Re \) and \( \Im \) denote the real and imaginary parts.
The first- and second-order moments \( \vec{d}_{\text{In}} \) and \( \sigma_{\text{In}} \) of a Gaussian input evolve through the sensor as
\begin{equation}
\begin{aligned}
	\vec{d}_{\text{Out}} &=  \vec{d}_{\text{In}} + \vec{d}_{\mathcal{V}}, &
	\sigma_{\text{Out}} &= \mathcal{S}_{\mathcal{M}} \sigma_{\text{In}} \mathcal{S}_{\mathcal{M}}^T.
\end{aligned}
\label{eq:moments_transformed}
\end{equation}

\textit{Quantum estimation.---}%
The \ac{CRB} and \ac{QCRB} are successive lower bounds on the variance \( (\Delta \estimator{h})^2 = \expectation{\estimator{h}^2} - \expectation{\estimator{h}}^2 \) of an unbiased estimator \( \estimator{h} \) for a parameter \( h \) which parameterises some probability distribution \( P(\vec{x}|h) \) and in turn some state \( \rho(h) \) which is given by
\begin{equation}
	(\Delta \estimator{h})^2
	\geq
	\frac{ 1 }{ F( h ) }
	\geq
	\frac{ 1 }{ H( h ) },
	\label{eq:crb}
\end{equation}
where \( F( h ) \) is the \ac{CFI} and \( H ( h ) \) the \ac{QFI}.
The \ac{CFI} depends on the sampled probability distribution as~\cite{braunstein_statistical_1994,hayashi_asymptotic_2005,paris_quantum_2009,toth_quantum_2014,demkowicz-dobrzanski_quantum_2015}
\begin{equation}
	F( h )
	=
	\sum\limits_{ \{ \vec{x} \} } \frac{1}{ P( \vec{x} | h ) } \left( \frac{ \partial P( \vec{x} | h ) }{ \partial h } \right)^2,
	\label{eq:fi}
\end{equation}
and the \ac{QFI} can be derived from the fidelity as~\cite{braunstein_statistical_1994,hayashi_asymptotic_2005,paris_quantum_2009,toth_quantum_2014,demkowicz-dobrzanski_quantum_2015}
\begin{equation}
	H ( h )
	=
	-4 \lim_{ \text{d}h \to 0 } \left\{
		\frac{\partial^2}{\partial(\text{d}h)^2} \mathcal{F}(\rho_{h},\rho_{h + \mathrm{d}h})
	\right\},
	\label{eq:qfi}
\end{equation}
where the fidelity is \( \mathcal{F}(\rho_1,\rho_2) = \trace{\sqrt{\sqrt{\rho_1}\rho_2\sqrt{\rho_1}}} \).
For single-parameter estimation there always exists some \ac{POVM} for which the second inequality of Eq.~\eqref{eq:crb} is saturated~\cite{braunstein_statistical_1994,paris_quantum_2009}.

For a parameter encoded only in the displacements of a Gaussian state the \ac{QFI} is~\cite{pinel_ultimate_2012,monras_phase_2013,gao_bounds_2014,safranek_quantum_2015}
\begin{equation}
	H( h )
	=
	2(\partial_{h} \vec{d})^T \sigma^{-1} (\partial_{h} \vec{d}),
	\label{eq:displacement_qcrb}
\end{equation}
where \( \vec{d} \) and \( \sigma \) are the displacement vector and covariance matrix of the Gaussian state respectively.

\( \mathcal{M} \) and \( \vec{\mathcal{V}} \) can be expressed as \( \mathcal{M} = B M B \) and \( \vec{\mathcal{V}} = B \vec{V} \),
where
\(
	B = \diag \left( e^{i\beta_1} \mathbb{1}_{2 \times 2}, \cdots , e^{i\beta_d} \mathbb{1}_{2 \times 2} \right)
\),
and \( M \) and \( \vec{V} \) are real for all cases given by Eq.~\eqref{eq:multiMV_inout}.
With an input state that can be written as \( \sigma_{\text{In}} = \sigma_0 \oplus \sigma_0 \), the \ac{QFI} for the parameter \( |h| \) is then (see Sec.~III of the supplementary material~\cite{note1}) given by
\begin{equation}
	H(|h|) = 4\vec{V}^T \left( M \sigma_0 M^T \right)^{-1} \vec{V}.
	\label{eq:gw_qfi}
\end{equation}
As Eq.~\eqref{eq:gw_qfi} is independent of \( \arg (h) \) we henceforth take \( h \) to be real and positive.

To compare with the spectral noise density which is typically used to describe the sensitivity of sensors~\cite{braginsky_quantum_1992,kimble_conversion_2001,miao_quantum_2014} the \acp{CRB} should be multiplied by \( 4 \) as \( S_n (\Omega) = 4/F(h) \), see Sec.~IV of the supplementary material~\cite{note1}.
Our bounds therefore have a pre-factor \( h_{\text{SQL}}^2 / 8 \) in comparison to results using the single-sided spectral density where the equivalent pre-factor is \( h_{\text{SQL}}^2 / 2 \)~\cite{kimble_conversion_2001,miao_quantum_2014}.

\textit{Sensor scheme.---}%
From Eq.~\eqref{eq:multiMV_inout} the optical modes are coupled through a multi-mode squeezing, which are weighted through the optical intensities of each mode.
We model optical loss at the detector by mixing the outgoing modes \( \opvec{b} \) with a (Gaussian) environment at a beam splitter with transmittivity \( \eta \) as
\(
	\opvec{b} \to \sqrt{\eta} \opvec{c} + \sqrt{1-\eta} \opvec{n},
\)
with reflected light dumped in a set of modes \( \opvec{n} \) which are traced out from the final state leaving only the measurable modes \( \opvec{c} \) accessible.
The effect on the final state is
\[
	\begin{aligned}
		\vec{d} \to \sqrt{ \eta } \vec{d}, && \sigma \to \eta \sigma + ( 1 - \eta ) \sigma_{\text{Loss}},
	\end{aligned}
\]
where we will take the input from the environment to be pure vacumm, namely \( \sigma_{\text{Loss}} = \mathbb{1} \).

Externally squeezed light inputs can enhance precision~\cite{caves_quantum-mechanical_1981,kimble_conversion_2001,the_ligo_scientific_collaboration_gravitational_2011} and has already been demonstrated in current \ac{GW} detectors~\cite{the_ligo_scientific_collaboration_gravitational_2011,the_ligo_scientific_collaboration_enhanced_2013}.
With multi-mode interferometers one feasible generalisation is to have parallel squeezing for the sidebands of each carrier frequency, with some squeezing \( \xi_j = r_j e^{i\phi_j} \) in the \( \op{x}_1^{(j)} \) and \( \op{x}_2^{(j)} \) modes.

Our main result is the fundamental quantum limit to the precision of the interferometer scheme described---with arbitrary intensity and external squeezing in each mode---which is
\begin{equation}
	\left( \Delta h \right)^2
	\geq
	\frac{h_{\text{SQL}}^2}{8}
	\left\{
	\frac{
		\left[1 -
		(1-\eta)\eta
	\expect{S \Gamma}\right]^2
		}{\eta\left[(1-\eta)\expect{\Gamma} + \eta \expect{Q \Gamma}\right]}
		+ (1-\eta)  \expect{P\Gamma}
\right\},
	\label{eq:qcrb_squeezeloss}
\end{equation}
where we define the diagonal matrices \( Q_{ii} = \left( \cosh 2r_i + \sinh 2r_i \cos 2\phi_i \right) \), \( S_{ii} = \sinh 2r_i \sin 2\phi_i \), \( \Gamma_{ii} = \left\{\left[(1-\eta)^2 + \eta^2\right]+ 2 \eta(1-\eta)\cosh 2r_i\right\}^{-1} \),
\( P = \eta\identity+(1-\eta)Q \), and \( \expect{A} \) is defined as \( \sum\limits_{i,j=1}^d \sqrt{\kappa_i \kappa_j} A_{ij} \).
The dependency on carrier mode intensity is a function of summations over \( \kappa_i \) weighted by various functions of the squeezing magnitude and angle in the respective mode. 

Attainability of quantum-limited precision requires the application of specific measurement schemes on the quantum system.
Homodyne detection covers both measurement of the signal quadrature which is in active use~\cite{kimble_conversion_2001,hild_dc-readout_2009,fricke_dc_2012,danilishin_quantum_2012} and the more general frequency-dependent homodyne~\cite{kimble_conversion_2001,danilishin_quantum_2012,miao_quantum_2014,mcclelland_lsc_2017} that measures along a different quadrature for each frequency mode \( \Omega \) of the signal. 
Both of these can be modelled by performing homodyne detection on some quadrature \( \sin\theta_i \op{x}^{(i)}_1 + \cos\theta_i \op{x}^{(i)}_2 \) for each carrier mode, in which $\theta_i$ can be frequency dependent.
This provides a precision of
\begin{align}
	\left( \Delta h \right)^2
	\geq 
	\frac{h_{\text{SQL}}^2}{8} 
	\Bigg\{ & \frac{\left[ 1-\eta\left( \expect{G^2 Y^{-1} S} + \expect{F G Y^{-1} Q} \right) \right]^2}{\eta\expect{G^2 Y^{-1}}} \nonumber \\
	&+  (1-\eta) \expect{Q Y^{-1}} + \eta \expect{G^2 Y^{-1}} \Bigg\},
	\label{eq:crb_squeezeloss}
\end{align}
where we further define the diagonal matrices \( F_{ii} = \sin \theta_i \), \( G_{ii} = \cos \theta_i \), and \( Y_{ii} = 1 - \eta + \eta[\cosh 2r_i - \sinh 2r_i \cos (2\phi_i + 2\theta_i)] \).

For measurements along the signal quadrature, \( F = 0 \), \( G = \mathbb{1} \), \( Y = T \) in Eq.~\eqref{eq:crb_squeezeloss} and the precision reduces to
\begin{equation}
	\left( \Delta h \right)^2
	\geq
	\frac{h_{\text{SQL}}^2}{8}
	\left[
		\frac{\left(1-\eta\expect{S T^{-1}}\right)^2}{\eta\expect{T^{-1}}}
		+
		\expect{P T^{-1}}
	\right],
	\label{eq:signalhom_crb}
\end{equation}
where \( T \) is the diagonal matrix \( T_{ii} = 1-\eta + \eta \left( \cosh 2r_i - \sinh 2r_i \cos 2\phi_i \right) \).

The bounds in Eqns.~\eqref{eq:qcrb_squeezeloss}--\eqref{eq:signalhom_crb} all take the form
\begin{equation}
	\frac{\Big( 1 - \sum\limits_i c_1^{(i)} \kappa_i \Big)^2}{\sum\limits_i c_2^{(i)}\kappa_i} + \sum\limits_i c_3^{(i)} \kappa_i,
	\label{eq:abstract_bound}
\end{equation}
for any given input squeezing configuration, with \( c_2^{(i)} > 0 \) and \( c_3^{(i)} \geq 0 \) with the equality \( c_3^{(i)} = 0 \) only holding if \( \eta = 1 \) which we consider explicitly as a special case later.
When \( c_2^{(i)} > 0 \) and \( c_3^{(i)} > 0 \), namely for \( \eta < 1 \), Eq.~\eqref{eq:abstract_bound} is always minimised (though not necessarily uniquely) over \( \{\kappa_i\} \in [0,\infty] \) by some \( \kappa_j = 1/\sqrt{(c_1^{(j)})^2 + c_2^{(j)}c_3^{(j)}} \) and \( \kappa_k = 0, \forall k \neq j \).
See Sec.~VII of the supplementary material~\cite{note1} for the complete proof.
This establishes our main conclusion that multi-carrier optomechanical sensors are fundamentally no better than their single carrier counterparts.

\textit{Special cases.---}%
With an identical external squeezing of \( r e^{i\phi} \) in each mode, the fundamental quantum limit in Eq.~\eqref{eq:qcrb_squeezeloss} becomes
\begin{widetext}
\begin{equation}
	\left( \Delta h \right)^2
	\geq
	\frac{h_{\text{SQL}}^2}{8}
	\frac{
		\eta^2
		+
		(1-\eta)
		\left[
		\left( 1-\eta \right) + 2\eta \cosh 2r
		-2\eta\kappa_{\text{Tot}} \sinh 2r\sin 2\phi
		+\eta \kappa_{\text{Tot}}^2 \left( \cosh 2r + \sinh 2r \sin 2\phi \right)
	\right]
}{\eta\kappa_{\text{Tot}} \left[(1-\eta) + \eta \left( \cosh 2r + \sinh 2r \sin 2\phi \right) \right]},
\label{eq:qcrb_equalsqueezing}
\end{equation}
where
\(
	\kappa_{\text{Tot}} = \sum\limits_{i=1}^d \kappa_i 
\) is the sole \( \kappa \)-dependent term.
In this case the fundamental quantum limit given in Eq.~\eqref{eq:qcrb_equalsqueezing} can be saturated with frequency-dependent homodyne using a homodyne angle of
\begin{equation}
	\theta_i =
	\arctan \left(
		\eta
		\frac{\kappa_{\text{Tot}}\left( \cosh 2r + \sinh 2r \cosh 2\phi \right) - \sinh 2r \sin 2\phi}{1-\eta+\eta \left( \cosh 2r + \sinh 2r \cos 2\phi \right)} 
	\right),  \forall i.
	\label{eq:hom_angle_equal}
\end{equation}

Measurement along the signal quadrature in this identical squeezing regime yields a precision of
\begin{equation}
	\left( \Delta h \right)^2
	\geq
	\frac{h_{\text{SQL}}^2}{8}
	\Bigg[
		\frac{1-\eta + \eta \left( \cosh 2r - \sinh 2r \cos 2\phi \right)}{\eta \kappa_{\text{Tot}}}
		+
		\kappa_{\text{Tot}} \left( \cosh 2r + \sinh 2r \cos 2\phi \right)
		-2\sinh 2r\sin 2\phi
	\Bigg],
	\label{eq:signalhom_equalsqueezing}
\end{equation}
\end{widetext}
which can be optimised by a frequency-dependent squeezing angle
\(
	\phi = \arctan \kappa_{\text{Tot}}
\),
to give a precision
\begin{equation}
	\left( \Delta h \right)^2
	\geq
	\frac{h_{\text{SQL}}^2}{8}
	\left[ 
		\frac{1-\eta + \eta e^{-2r}}{\eta \kappa_{\text{Tot}}} + e^{-2r}\kappa_{\text{Tot}}
	\right].
	\label{eq:signalhom_equalsqueezing_optimalangles}
\end{equation}

In the limit of zero squeezing, the fundamental quantum limit reduces to
\begin{equation}
	\left( \Delta h \right)^2
	\geq
	\frac{h_{\text{SQL}}^2}{8}
	\left[
		\frac{1}{\eta \kappa_{\text{Tot}}} + (1-\eta) \kappa_{\text{Tot}}
	\right].
	\label{eq:qcrb_unsqueezed}
\end{equation}
This takes the same form as the single-mode limit~\cite{kimble_conversion_2001,miao_quantum_2014} with \( \kappa_{\text{Tot}} \) taking the place of the single carrier \( \kappa \).

Using the frequency-dependent homodyne angle given by Eq.~\eqref{eq:hom_angle_equal}, this precision can be attained with the homodyne angle \( \arctan \left( \eta \kappa_{\text{Tot}} \right) \).
Considering homodyne along the signal quadrature instead, the precision given by Eq.~\eqref{eq:signalhom_crb} reduces to
\begin{equation}
	\left( \Delta h \right)^2
	\geq
	\frac{h_{\text{SQL}}^2}{8}
	\left[
		\frac{1}{\eta \kappa_{\text{Tot}}} + \kappa_{\text{Tot}}
	\right].
	\label{eq:signalhom_unsqueezed}
\end{equation}

In the lossless limit \( \eta = 1 \) with squeezings not necessarily identical across the carriers,  the fundamental quantum limit is
\begin{equation}
	\left( \Delta h \right)^2
	\geq
	\frac{ h_{\text{SQL}}^2 }{8}
	\frac{1}{ K_{\text{Tot}} },
	\label{eq:qcrb_losslesssqueeze}
\end{equation}
where we define \( K_{\text{Tot}} \) as \( K_{\text{Tot}} = \sum\limits_{i=1}^d \kappa_i \left( \cosh 2r_i +\sinh 2r_i \cos 2\phi_i \right) \), annd the bound displays shot-noise behaviour, being minimised as \( \kappa_i \to \infty \).
This bound is attained by the frequency-dependent homodyne angle
\begin{equation}
	\theta_i = \arctan
	\left( \frac{K_{\text{Tot}} - \sinh 2r_i \sin 2\phi_i}{\cosh 2r_i + \sinh 2r_i \cos 2\phi_i} \right), \forall i,
	\label{eq:hom_angle_lossless}
\end{equation}
while a squeezing angle \( \phi_i = 0, \forall i \) optimises the precision.

\textit{Conclusions and discussions.---}%
We have shown that no improvement is afforded in the fundamental sensitivity bound in a large class of optomechanical sensors by simultaneous use of multiple carrier modes, including under the effect of optical loss.
With identical squeezing in each mode the precision is determined solely by \( \kappa_{\text{Tot}} \) and no other properties of the distribution of \( \{ \kappa_i \} \).
Introducing squeezing with different magnitudes of angles breaks this symmetry but the optimum interferometer configuration is not enhanced by the presence of multiple carriers.

\textit{Acknowledgements.---}We would like to thank members of the 
LSC AIC, MQM, and QN groups for fruitful discussions. 
D.B.\ and A.D.\ are supported, in part, by the UK EPSRC (EP/K04057X/2), and the UK National Quantum Technologies Programme (EP/M01326X/1, EP/M013243/1).
H.M.\ is supported by UK STFC Ernest Rutherford 
Fellowship (Grant No.\ ST/M005844/11). 

\bibliography{tuned_multimode}

\begin{thebibliography}{11}
\bibitem [{Note1()}]{note1}%
  \BibitemOpen
  \bibinfo {note} {Values used are \( I_{\protect \text {Tot}} = \SI {840}{\kW
  } \), \( L = \SI {4}{\km } \), \( M = \SI {40}{\kg } \), \( \gamma _i \approx
  \gamma = \SI [product-units=single]{2\pi x 500}{\per \s } \), \( \omega _i
  \approx \omega = \SI [product-units=single]{2\pi x 2.82e14}{\per \s }
  \)}\BibitemShut {NoStop}%
\end{thebibliography}

\end{document}


\title{Supplemental Material: Fundamental Quantum Limits of Multicarrier Optomechanical Sensors}

\author{Dominic Branford}
\affiliation{Department of Physics, University of Warwick, Coventry CV4 7AL, United Kingdom}

\author{Haixing Miao}
\affiliation{School of Physics and Astronomy, Institute of Gravitational Wave Astronomy, University of Birmingham, Birmingham B15 2TT, United Kingdom}

\author{Animesh Datta}
\affiliation{Department of Physics, University of Warwick, Coventry CV4 7AL, United Kingdom}

\date{\today}

\maketitle

\setcounter{secnumdepth}{3}

\onecolumngrid

\section{Phase space with respect to the two-photon formalism}
\label{app:twophoton_gaussian}
From the two-photon operators
\begin{align}
	\op{a}_1 &= \frac{ \op{a}_{ \omega + \Omega } + \op{a}_{ \omega - \Omega }^{\dagger} }{ \sqrt{2} },
	&
	\op{a}_2 &= \frac{ \op{a}_{ \omega + \Omega } - \op{a}_{ \omega - \Omega }^{\dagger} }{ i\sqrt{2} },
\end{align}
one can construct hermitian position and momentum operators from \( \op{a}_1 \) and \( \op{a}_2 \) as
\begin{align}
	\op{x}_1 &= \frac{ \op{a}_1 + \op{a}_1^{\dagger} }{ \sqrt{2} } = \frac{ \op{a}_{+} + \op{a}_{-}^{\dagger} + \op{a}_{+}^{\dagger} + \op{a}_{-} }{ 2 } = \frac{ \op{x}_{+} + \op{x}_{-} }{ \sqrt{2} }, \\
	\op{x}_2 &= \frac{ \op{a}_2 + \op{a}_2^{\dagger} }{ \sqrt{2} } = \frac{ \op{a}_{+} - \op{a}_{-}^{\dagger} - \op{a}_{+}^{\dagger} + \op{a}_{-} }{ 2i } = \frac{ \op{p}_{+} + \op{p}_{-} }{ \sqrt{2} }, \\
	\op{p}_1 &= \frac{ \op{a}_1 - \op{a}_1^{\dagger} }{ i\sqrt{2} } = \frac{ \op{a}_{+} + \op{a}_{-}^{\dagger} - \op{a}_{+}^{\dagger} - \op{a}_{-} }{ 2i } = \frac{ \op{p}_{+} - \op{p}_{-} }{ \sqrt{2} }, \\
	\op{p}_2 &= \frac{ \op{a}_2 - \op{a}_2^{\dagger} }{ i\sqrt{2} } = \frac{ - \op{a}_{+} + \op{a}_{-}^{\dagger} - \op{a}_{+}^{\dagger} + \op{a}_{-} }{ 2 } = \frac{ - \op{x}_{+} + \op{x}_{-} }{ \sqrt{2} },
\end{align}
where we adopt \( \op{a}_{\pm} \) as shorthand for \( \op{a}_{ \omega \pm \Omega } \), \( \op{x}_{\pm} \) for \( \op{x}_{ \omega \pm \Omega } \), and \( \op{p}_{\pm} \) for \( \op{p}_{ \omega \pm \Omega } \).
We can recognise from this that the position and momentum operators of the two-photon creation/annihilation operators correspond to a rotation of the sideband frequency creation/annihilation operators.
Thus we shall express states as a function of the \( \{ \op{x}_1, \op{x}_2, \op{p}_1, \op{p}_2 \} \) operators which we see is equivalent to the use of the \( \{ \op{x}_+, \op{p}_+, \op{x}_-, \op{p}_- \} \) operators through a simple change of basis.
Here the non-zero commutators,
\(
	\commutator{ \op{x}_+(\Omega_A) }{ \op{p}_+(\Omega_B) } = \commutator{ \op{x}_-(\Omega_A) }{ \op{p}_-(\Omega_B) } = i\delta(\Omega_A-\Omega_B) 
\)
in terms of the frequency quadratures, have become
\(
	\commutator{ \op{x}_1(\Omega_A) }{ \op{x}_2(\Omega_B) } = \commutator{ \op{p}_1(\Omega_A) }{ \op{p}_2(\Omega_B) } = i\delta(\Omega_A-\Omega_B)
\).

\section{Input-output relations for a gravitational-wave interferometer}
Optical interferometers typically require mirrors to redirect light back to a common point in order to generate interference.
When the motion of the mirrors are disturbed by the reflected light a squeezing of the optical fields is produced through the interaction between optical and mechanical modes.
This optomechanical effect is particularly apparent in systems used to resolve the displacement of the mechanical system such as laser-interferometric \ac{GW} detectors.

For the tuned interferometer optical fields evolve through the interferometer as~\cite{miao_quantum_2012} 
\begin{align}
	\op{b}_1^{(j)}(t) &= \op{a}_1^{(j)}(t-2\tau),\label{eq:b1time}\\
	\op{b}_2^{(j)}(t) &= \op{a}_2^{(j)}(t-2\tau) + \sqrt{\frac{2I_j}{\hbar\omega_j}}\frac{\omega_j}{c} \op{x}_d(t-\tau),\label{eq:b2time}
\end{align}
where \( \op{x}_d \) is the differential motion of the two mirrors and common to all optical modes.
The signal to sense \( h(t) \) acts on the mechanical part of the optomechanical sensor as a force causing the interferometer arm lengths to vary.
In the tuned configuration differential motion of the mirrors is thus
\begin{equation}
	m \ddot{\op{x}}_d (t) + m \Omega_p \op{x}_d (t) = 4 \sum\limits_{i} \sqrt{\frac{2\hbar\omega_iI_i}{c^2}} \op{a}_1^{(i)}(t-\tau) + mL\ddot{h}(t).
	\label{eq:xd_time}
\end{equation}
Translating the optical field evolution to the frequency domain Eqs.~\eqref{eq:b1time} and~\eqref{eq:b2time} become
\begin{align}
	\op{b}_1^{(j)}(\Omega) &= e^{2i\Omega\tau_j} \op{a}_1^{(j)}(\Omega), \\
	\op{b}_2^{(j)}(\Omega) &= e^{2i\Omega\tau_j} \op{a}_2^{(j)}(\Omega) - \frac{e^{i\Omega\tau_j}}{m(\Omega^2 - \Omega_p^2)}\sqrt{\frac{2I_j}{\hbar\omega_j}}\frac{\omega_j}{c} \left[ \sum\limits_i 4e^{i\Omega\tau_i} \sqrt{\frac{2\hbar\omega_iI_i}{c^2}} \op{a}_1^{(i)}(\Omega) - m L \Omega^2 h(\Omega) \right],
	\label{eq:app:b2freq}
\end{align}
where the \( 1/\left( \Omega^2 - \Omega_p^2 \right) \) term produces a resonance at \( \Omega = \Omega_p \).

From this we derive the expressions for multi-mode input-output relations (Eq.~(2) in the main text)
\begin{equation}
\begin{aligned}
	\mathcal{M}_{jk} &= 
	e^{i(\beta_j+\beta_k)}
	\begin{pmatrix}
		\delta_{jk} & 0 \\
		- \chi\sqrt{ \kappa_j \kappa_k } & \delta_{jk}
	\end{pmatrix},
	&
	\mathcal{V}_{j} &= 
	\chi
	\frac{ e^{i\beta_j} }{ h_{\text{SQL}} }
	\begin{pmatrix}
		0 \\
		\sqrt{ 2 \kappa_j }
	\end{pmatrix}.
\end{aligned}
\label{eq:app:multiMV_inout}
\end{equation}
where
\begin{equation}
	\begin{aligned}
		\chi &=  \sign \left( \Omega^2 - \Omega_p^2 \right) &
		\kappa_i &= \left| \frac{2\sqrt{2} I_i \omega_i}{ m c^2 \left( \Omega^2 - \Omega_p^2 \right) } \right| &
		\beta_i &= \Omega \tau_i &
		h_{\text{SQL}} &= \sqrt{\frac{4\hbar}{m L^2 \Omega^2}}
	\end{aligned}
\end{equation}
give the input-output relations for a simplified interferometer, accounting for the cavity modes retrieves the form given in Eq.~(3) in the main text~\cite{miao_quantum_2012}.
In the tuned configuration the pendulum response can typically be found around \( \SI{\sim1}{\Hz} \)~\cite{corbitt_squeezed-state_2006,the_ligo_scientific_collaboration_advanced_2015} which lies below the principal frequency range of advanced LIGO~\cite{the_ligo_scientific_collaboration_advanced_2015} and so it is sufficient to take \( \Omega_p \ll \Omega \) and consider \( \chi=1 \) and drop the \( \Omega_p \) from the definition of \( \kappa_i \).

Through the signal-recycling mirror~\cite{buonanno_quantum_2001,buonanno_signal_2002} the mechanical response is modified to exhibit a spring-like reaction at some higher frequency \( \Theta \) which does appear at larger frequencies.
While the input-output relations are generally more complicated than those of Eq.~\eqref{eq:app:multiMV_inout} in the low-frequency regime---where radiation pressure dominates---the input-output relations can be reduced to follow the form of Eq.~\eqref{eq:app:multiMV_inout}~\cite{corbitt_squeezed-state_2006}.
In the low-frequency domain, with mirror motion well below the cavity bandwidth (\( \Omega \ll \gamma_i \)) the signal-recycling mirror configuration reduces to the same form of input-output relations as Eq.~\eqref{eq:app:multiMV_inout} with the optomechanical couplings \( \kappa_i \) being proportional to \( 1/\left( \Omega^2 - \Theta^2 \right) \) and \( \chi = \sign (\Omega^2 - \Theta^2) \).

The two cases \( \chi = 1 \) and \( \chi = -1 \) can be related through the phase shift \( D = \bigoplus\limits_{i=1}^d \begin{pmatrix} 1 & 0 \\ 0 & \chi \end{pmatrix} \) and so the \( \chi = 1 \) response is equivalent to an interferometer with negative response which undergoes a \( \pi \) phase shift acting on \( \opvec{x}_2 \) and \( \opvec{p}_2 \) before the initial state is input to the sensor and after the state is output from the sensor.

Specifically for a squeezed vacuum input such as in Eq.~\eqref{eq:app:squeezed_input}
\[
	\sigma = \bigoplus_{i=1}^2 \left[ \bigoplus_{j=1}^d \begin{pmatrix} \cosh 2r_j + \sinh 2r_j \cos 2\phi_j & \sinh 2r_j \sin 2\phi_j \\ \sinh 2r_j \sin 2\phi_j & \cosh 2r_j - \sinh 2r_j \cos 2\phi_j \end{pmatrix} \right],
\]
if then followed by a phase shift \( D \) which acts as \( (D \oplus D) \sigma (D \oplus D) \) leaving the squeezing maginitudes unchanged and negates the squeezing angles
\[
	\begin{aligned} (D \oplus D) & \sigma(r_1,\phi_1,\cdots,r_d,\phi_d) (D \oplus D) \\
	&=  (D \oplus D) \left\{ \bigoplus_{i=1}^2 \left[ \bigoplus_{j=1}^d \begin{pmatrix} \cosh 2r_j + \sinh 2r_j \cos 2\phi_j & -\sinh 2r_j \sin 2\phi_j \\ -\sinh 2r_j \sin 2\phi_j & \cosh 2r_j - \sinh 2r_j \cos 2\phi_j \end{pmatrix} \right] \right\} (D \oplus D) \\
	&= \sigma(r_1,-\phi_1,\cdots,r_d,-\phi_d),
\end{aligned}
\]
the squeezing angles \( \{ \phi_i \} \) when \( \chi = -1 \) give the same sensitivity as the squeezing angles \( \{ -\phi_i \} \) in the absence of the phase shift.
Similarly performing a rotation \( \theta_i \) between \( \op{x}_1^{(i)} \) and \( \op{x}_2^{(i)} \) as done to model homodyne detection in Sec.~\ref{app:hom_squeezeloss} after the phase operation \( D \) is equivalent to performing the rotation \( -\theta_i \) without the phase shift.
This allows us to recover sensitivities for the \( \chi = -1 \) response from the expressions for the \( \chi = 1 \) response and apply our conclusions to either sign of the response.

\section{Quantum Cramér-Rao bound for estimating a displacement from input-output relations}
\label{app:inout_qcrb}
For a Gaussian input state whose covariance matrix is block-diagonal, namely \( \sigma_{\text{dark}} = \sigma_0 \oplus \sigma_0 \), and has an input-output relation of the form
\begin{equation}
	\opvec{b}(\Omega) = B M B \opvec{a}(\Omega) + h(\Omega) B \vec{V}(\Omega),
\end{equation}
where \( M \in \mathbb{R}^{2d{\times}2d} \) and \( \vec{V} \in \mathbb{R}^{2d} \), and
\(
	B = \diag ( e^{i\beta_1}, e^{i\beta_1}, e^{i\beta_2}, e^{i\beta_2}, \cdots, e^{i\beta_d}, e^{i\beta_d} )
\).

The evolved state is given by
\begin{equation}
\begin{aligned}
	\vec{d} &= \sqrt{2} \mathcal{S}_B \begin{pmatrix} \Re[h] \vec{V} \\ \Im[h] \vec{V} \end{pmatrix},
	&
	\sigma &=
	\mathcal{S}_B
	\begin{pmatrix}
	M \left( \Re[B] \sigma_0 \Re[B] + \Im[B] \sigma_0 \Im[B] \right) M^T & 0 \\
	0 & M \left( \Re[B] \sigma_0 \Re[B] + \Im[B] \sigma_0 \Im[B] \right) M^T
	\end{pmatrix}
	\mathcal{S}_B^T.
\end{aligned}
\label{eq:app:ideal_moments}
\end{equation}
The \ac{QFI} for the magnitude of any signal \( |h| = \sqrt{( \Re h )^2 + ( \Im h )^2}\) is given by
\begin{equation}
	H(|h|) = 2 ( \partial_{|h|} \vec{d} )^T \sigma^{-1} ( \partial_{|h|} \vec{d} ),
\end{equation}
which for \( \vec{d} \) and \( \sigma \) of form
\begin{equation}
	\begin{aligned}
		\vec{d} &= \mathcal{S}\begin{pmatrix} \Re[h] \vec{W} \\ \Im[h] \vec{W} \end{pmatrix}, &
		\sigma &= \mathcal{S}\begin{pmatrix} \sigma_N & 0 \\ 0 & \sigma_N \end{pmatrix}\mathcal{S}^T,
	\end{aligned}
	\label{eq:app:abstract_moments}
\end{equation}
is
\begin{equation}
	H(|h|) = 2\left[ ( \partial_{|h|} \Re h )^2 + ( \partial_{|h|} \Im h )^2 \right] \vec{W}^T \sigma_N^{-1} \vec{W},
\end{equation}
as \( ( \partial_{|h|} \Re h )^2 + ( \partial_{|h|} \Im h )^2 = 1 \) the \ac{QFI} can thus be reduced to
\begin{equation}
	H(|h|) = 2\vec{W}^T \sigma_N^{-1} \vec{W},
\end{equation}
which is equivalent to the \ac{QFI} obtained by taking the signal to be real.

The ideal state in Eq.~\eqref{eq:app:ideal_moments} is of the form of Eq.~\eqref{eq:app:abstract_moments} and remains such under mixture with any Gaussian state of the same form, including thermal states which have a diagonal covariance matrix, producing a state \( \eta \left( \Re[B] \sigma_0 \Re[B] + \Im[B] \sigma_0 \Im[B] \right) + (1-\eta)\sigma_1 \).
\begin{equation}
	H(|h|) = 4 \vec{V}^T
	\left[
		\eta M \left( \Re[B] \sigma_0 \Re[B] + \Im[B] \sigma_0 \Im[B] \right) M^T + (1-\eta) \sigma_1
	\right]^{-1}
	\vec{V},
	\label{eq:app:qfi_magnitude}
\end{equation}
Thus taking \( h \in \mathbb{R} \) allows us to consider only the \( \opvec{x} \) modes as the state has a covariance matrix with block form \( \sigma_0 \oplus \sigma_0 \) it is separable between the \( \opvec{x} \) and \( \opvec{p} \) modes; with the latter modes contain no parameter dependence and being uncorrelated with any of the modes which have a parameter-dependence the \( \opvec{p} \) modes can be discarded from our analysis.

If the input covariance matrix \( \sigma_0 \) and the covariance matrix the state is mixed with at the output \( \sigma_1 \) are block-diagonal with block size \( n \) and \( B \) is of the form \( \bigoplus_j e^{i\beta_j} \identity_{n {\times} n } \), then this simplifies through \( \Re[B] \sigma_0 \Re[B] + \Im[B] \sigma_0 \Im[B] = \sigma_0 \) and \( \Re[B] \sigma_1 \Re[B] + \Im[B] \sigma_1 \Im[B] = \sigma_1 \).
\begin{equation}
	H(|h|) = 4 \vec{V}^T \left( \eta M \sigma_0 M^T + (1-\eta) \sigma_1 \right)^{-1} \vec{V}.
\end{equation}
This is the case for systems considered in Secs.~\ref{app:qcrb_squeezeloss} and~\ref{app:hom_squeezeloss} where the externally input squeezing is localised to a carrier and so \( \sigma_0 \) is block-diagonal with \( 2 {\times} 2 \) blocks.

\section{Relation between the Cramér-Rao bound and spectral density for a signal with white Gaussian noise}
\label{app:crb_spectraldensities}
For estimating a signal \( h(t) \) from a measured signal \( y(t) = h(t) + w(t) \), which is a stationary process, the sensitivity can be measured by the single-sided spectral density~\cite{braginsky_quantum_1992,thorne_modern_2017}
\begin{equation}
	S(\Omega) = 2 \int\limits_{-\infty}^{\infty} \intd \tau \, C_w(\tau) \cos(\Omega\tau),
\end{equation}
where \( C_w (\tau) \) is the autocorrelation function
\begin{equation}
	C_w (\tau) = \lim\limits_{T\to\infty} \frac{1}{2T} \int\limits_{-T}^{T} \intd t \, [w (t) - \bar{w}] [w (t+\tau) - \bar{w}],
\end{equation}
where \( \bar{w} \) is the time-average of \( w \).
For a white Gaussian noise process \( C_w(\tau) = \upsilon \delta (\tau) \), this is simply
\begin{equation}
	S(\Omega) = 2 \upsilon.
	\label{eq:app:spectral_gen}
\end{equation}

For the same signal \( y(t) = h(t) + w(t) \) the precision of any estimator of a parameter \( g \) of the signal is~\cite{kay_fundamentals_1998}
\begin{equation}
(\Delta h)^2 \geq
	\frac{\upsilon}{\mathbb{E} \left[ \left( \frac{ \partial h(t) }{ \partial g } \right)^2 \right]}.
\end{equation}
The equivalent case of interest to the spectral noise density is the amplitude of the frequency modes \( h(\Omega) \), for which the relevant derivative is
\begin{equation}
\begin{aligned}
	\frac{\partial h(t)}{\partial |h(\Omega)|} &=
	\frac{\partial}{\partial |h(\Omega)|}
	\int\limits_{-\infty}^{\infty} \intd \Omega' \, h(\Omega') e^{i\Omega' t} \\
	&=
	\int\limits_{-\infty}^{\infty} \intd \Omega' \, \left[ \delta(\Omega-\Omega') e^{i\arg (h(\Omega'))} + \delta(\Omega+\Omega') e^{i\arg (h(\Omega'))} \right] e^{i\Omega' t}\\
	&=
	2\cos \left[ \Omega t + \arg(h(\Omega)) \right],
\end{aligned}
\end{equation}
where the derivative yields two terms due to the Fourier transform property \( h(-\Omega) = h(\Omega)^{\dagger} \).
The expectation of the square of \( \partial_{|h(\Omega)|} h(t) \) is then simply \( 2 \) giving a \ac{CRB} of
\begin{equation}
	(\Delta |h(\Omega)|)^2 \geq \frac{\upsilon}{2},
	\label{eq:app:crb_gen}
\end{equation}
showing a proportionality constant of \( 4 \) relating these two methods of calculating sensitivities given by Eqs.~\eqref{eq:app:spectral_gen} and~\eqref{eq:app:crb_gen}.

\section{Quantum Cramér-Rao bound for a lossy interferometer with squeezed vacuum input}
\label{app:qcrb_squeezeloss}
Parallel squeezing corresponds to the covariance matrix of the \( \opvec{x} \) operators on the input dark port being
\begin{equation}
	\sigma_{\text{dark}}
	=
	\bigoplus_{i=1}^d
	\begin{pmatrix}
		\cosh 2r_i + \sinh  2r_i \cos 2\phi_i &
		\sinh 2r_i \sin 2\phi_i \\
		\sinh 2r_i \sin 2\phi_i &
		\cosh 2r_i - \sinh  2r_i \cos 2\phi_i
	\end{pmatrix},
	\label{eq:app:squeezed_input}
\end{equation}
which evolves through the interferometer to
\begin{equation}
	\sigma_{ij}
	=
	\begin{multlined}
	\left(
		\begin{array}{c}
		\delta_{ij} (\cosh 2r_i + \sinh 2r_i \cos 2\phi_i) \\
		\delta_{ij} \sinh 2r_i \sin 2\phi_i - \sqrt{\kappa_i \kappa_j}(\cosh 2r_j + \sinh 2r_j \cos 2\phi_j)
		\end{array}
	\right.
	\\
	\shoveright{
	\left.
		\begin{array}{c}
		\delta_{ij} \sinh 2r_i \sin 2\phi_i - \sqrt{\kappa_i \kappa_j}(\cosh 2r_i + \sinh 2r_i \cos 2\phi_i) \\
		\delta_{ij} (\cosh 2r_i - \sinh 2r_i \cos 2\phi_i) - \sqrt{\kappa_i \kappa_j} \left( \sinh 2r_i \sin 2\phi_i + \sinh 2r_j \sin 2\phi_j - K_{\text{Tot}} \right)
		\end{array}
	\right)
	}
	\end{multlined}
\end{equation}
where each mode has a squeezing \( \xi_k = r_k e^{i\phi_k} \), and
\[
	K_{\text{Tot}} = \sum\limits_{k=1}^d \kappa_k \left( \cosh 2r_k + \sinh 2r_k \cos 2\phi_k \right),
\]
represents a squeezed version of \( \kappa_{\text{Tot}} \).

The effect of loss on \( \sigma \) is to mix the matrix with the state \( \sigma_{\text{Loss}} \), which we will take to be the vacuum state, under
\[
	\sigma \to \eta \sigma + (1-\eta) \sigma_{\text{Loss}},
\]
produces the Gaussian state on the \( \opvec{x} \) modes
\begin{equation}
	\sigma_{ij}
	=
	\begin{multlined}[t]
	\delta_{ij}
\begin{pmatrix}
	\eta\left( \cosh 2r_i + \sinh 2r_i \cos 2\phi_i \right) + (1-\eta) &
	\eta\sinh 2r_i \sin 2\phi_i \\
	\eta\sinh 2r_i \sin 2\phi_i &
	\eta\left( \cosh 2r_i - \sinh 2r_i \cos 2\phi_i \right) + (1-\eta)
\end{pmatrix}\\
-\eta\sqrt{\kappa_i \kappa_j}
\begin{pmatrix}
	0 &
	\cosh 2r_i + \sinh 2r_i \cos 2\phi_i \\
	\cosh 2r_j + \sinh 2r_j \cos 2\phi_j &
	\sinh 2r_i \sin 2\phi_i + \sinh 2r_j \sin 2\phi_j - K_{\text{Tot}}
\end{pmatrix}.
\end{multlined}
\label{eq:app:cov_squeezeloss}
\end{equation}
This covariance matrix has no obvious inverse, however we can rearrange \( \sigma \) in block form such that the top left quarter is the covariances of the \( \opvec{x}_1^{(\omega_i)} \) operators and observe that this can be rewritten in terms of the matrices \( Q \), \( R \), \( S \), and \( L \) where
\begin{equation}
\begin{aligned}
	Q_{ij} &= \delta_{ij} \left( \cosh 2r_i + \sinh 2r_i \cos 2\phi_i \right),\\
	R_{ij} &= \delta_{ij} \left( \cosh 2r_i - \sinh 2r_i \cos 2\phi_i \right),\\
	S_{ij} &= \delta_{ij} \sinh 2r_i \sin 2\phi_i, \\
	L_{ij} &= \vec{k}\vec{k}^T = \sqrt{\kappa_i \kappa_j},
\end{aligned}
\end{equation}
such that
\begin{equation}
	\sigma
	=
	\begin{pmatrix}
		(1-\eta) \identity + \eta Q &
		\eta \left( S - Q L \right) \\
		\eta \left( S - L Q \right) &
		(1-\eta) \identity + \eta \left( R - S L - L S + K_{\text{Tot}} L \right)
	\end{pmatrix},
	\label{eq:app:cov_squeezeloss_rearranged}
\end{equation}
the inverse can then be found, with all blocks being invertible with the Woodbury matrix identity
\begin{equation}
	\left( A + U C V \right)^{-1}
	=
	A^{-1} - A^{-1} U \left( C^{-1} + V A^{-1} U \right)^{-1} V A^{-1},
	\label{eq:woodbury}
\end{equation}
The parameter information is encoded in the \( \{ \op{x}_2^{(\omega_i)} \} \) modes and so calculation of the \ac{QCRB} requires the lower right quarter of the inverse.
For a block matrix, the inverse is
\[
	\begin{pmatrix}
		A & B \\ C & D
	\end{pmatrix}^{-1}
	=
	\begin{pmatrix}
		\left( A - B D^{-1}C \right)^{-1} &
		- \left( A - B D^{-1}C \right)^{-1} B D^{-1} \\
		- D^{-1} C \left( A - B D^{-1}C \right)^{-1} &
		D^{-1} + D^{-1} C \left( A - B D^{-1}C \right)^{-1} B D^{-1}
	\end{pmatrix}.
\]
The relevant inverses are given---using Eq.~\eqref{eq:woodbury}---by
\begin{equation}
	D^{-1} =
	T^{-1}
	-
	\frac{\eta}{\alpha}
	T^{-1}
	\begin{pmatrix} \vec{k} & S\vec{k} \end{pmatrix}
		\begin{pmatrix}
			K_{\text{Tot}} - \eta\expect{ST^{-1}S} &
			-1 + \eta\expect{T^{-1}S} \\
			-1 + \eta\expect{T^{-1}S} &
			-\eta\expect{T^{-1}}
		\end{pmatrix}
	\begin{pmatrix} \vec{k}^T \\ \vec{k}^T S \end{pmatrix}
	T^{-1},
\end{equation}
where we introduce the definitions \( \expect{Z} = \trace{L Z} \), and can thus rewrite \( K_{\text{Tot}} \) as \( \expect{Q} \), and  define \( T = (1-\eta) \identity + \eta R \) and \( \alpha = 1 + \eta\left[ K_{\text{Tot}}\expect{T^{-1}} - 2 \expect{S T^{-1} S}+ \eta \left( \expect{S T^{-1}}^2 - \expect{T^{-1}}\expect{S T^{-1} S} \right) \right] \).
As well as
\begin{equation}
	\left( A - B D^{-1} C \right)^{-1}
	=
	W^{-1}
	-
	W^{-1}
	\begin{pmatrix} S T^{-1} \vec{k} & U \vec{k} \end{pmatrix}
	X^{-1}
	\begin{pmatrix} \vec{k}^T T^{-1} S \\ \vec{k}^T U \end{pmatrix}
	W^{-1},
\end{equation}
where \( U = \left( Q - \eta S T^{-1} S \right) \), \( W = (1-\eta)\identity + \eta Q - \eta^2 S T^{-1} S \), and
\[
	X
	=
	\begin{pmatrix}
		\frac{\expect{T^{-1}}}{\eta^2} + \expect{T^{-1} S W^{-1} S T^{-1}} &
		\frac{1 - \eta \expect{S T^{-1}}}{\eta^2} + \expect{T^{-1} S W^{-1} U} \\
		\frac{1 - \eta \expect{S T^{-1}}}{\eta^2} + \expect{T^{-1} S W^{-1} U} &
		\frac{-\expect{Q} + \eta \expect{S T^{-1} S}}{\eta} + \expect{U W^{-1} U}
	\end{pmatrix}.
\]
From this the \ac{QFI} can be evaluated with
\[
	H(h) =
	\frac{8\eta}{h_{\text{SQL}}^2}
	\sum\limits_{i,j=1}^d
	\sqrt{\kappa_i \kappa_j} \left( \sigma^{-1} \right)_{d+i,d+j}.
\]
From these expressions we can (after much simplification) construct the \ac{QFI}
\begin{equation}
	H(h)
	=
	\frac{8\eta}{h_{\text{SQL}}^2}
	\frac{(1-\eta)\expect{\Gamma} + \eta \expect{Q \Gamma}}
	{
		1 -
		(1-\eta)\eta \left\{
			\expect{S \Gamma} \left[ 2 - (1-\eta)\eta\expect{S \Gamma} \right]
	- \left[ \eta \expect{\Gamma} + (1-\eta) \expect{Q \Gamma} \right] \left[ (1-\eta) \expect{\Gamma} + \eta \expect{Q \Gamma} \right] \right\}
},
\end{equation}
where
\[
	\Gamma = T^{-1} W^{-1} = \left\{\left[(1-\eta)^2 + \eta^2\right]\identity + \eta(1-\eta)\left( Q + R \right)\right\}^{-1},
\]
from which the \ac{QCRB} is given by
\begin{equation}
	\left( \Delta h \right)^2
	\geq
	\frac{h_{\text{SQL}}^2}{8}
	\Bigg\{
		\frac{1}{\eta(1-\eta)\expect{\Gamma} + \eta^2 \expect{Q \Gamma}}
			- (1-\eta) \Bigg[
			2\frac{\expect{S \Gamma}}{(1-\eta)\expect{\Gamma} + \eta \expect{Q \Gamma}}
			- \eta(1-\eta) \frac{\expect{S \Gamma}^2}{(1-\eta)\expect{\Gamma} + \eta \expect{Q \Gamma}} - \left(\eta \expect{\Gamma} + (1-\eta) \expect{Q \Gamma} \right)
	\Bigg]\Bigg\}.
	\label{eq:app:qcrb_loss_squeeze}
\end{equation}
If we assume an equal squeezing mode and angle in each mode this \ac{QCRB} reduces to
\begin{equation}
	\left( \Delta h \right)^2
	\geq
	\frac{h_{\text{SQL}}^2}{8}
	\left\{
		\frac{
			1+2\eta(1-\eta)\left( \cosh 2r - 1 \right) -2\eta(1-\eta)\kappa_{\text{Tot}}\sinh 2r \sin 2\phi + \eta(1-\eta)\kappa_{\text{Tot}}K_{\text{Tot}}
		}{\eta\left[ (1-\eta)\kappa_{\text{Tot}} + \eta K_{\text{Tot}} \right]}
	\right\}.
	\label{eq:app:qcrb_loss_equal_squeeze}
\end{equation}

\section{Cramér-Rao bounds for a lossy interferometer with squeezed vacuum input under homodyne detection}
\label{app:hom_squeezeloss}
The results of homodyne measurement are given by the marginal distribution of the Wigner function consisting of a set of commuting quadratures~\cite{adesso_continuous_2014}.
Homodyne measurement local to each carrier-mode would consist of performing homodyne read-out with some angle \( \theta_i \) between the two \( \{ \op{x}_1^{(\omega_i)} , \op{x}_2^{(\omega_i)} \} \) modes, where we omit the \( \opvec{p} \) modes by takin \( h \) to be real and thus leaving the statistics of these modes independent of \( h \).
Providing this angle is equivalent to performing a rotation \( \theta_i \) between the \( \op{x}_1^{(i)} \) and \( \op{x}_2^{(i)} \) modes
\begin{equation}
	\mathcal{S}_{\text{Hom.}}(\vec{\theta})
	=
	\begin{pmatrix}
		\cos \theta_1 & -\sin \theta_1 & \cdots & 0 & 0 \\
		\sin \theta_1 & \cos \theta_1 & \cdots & 0 & 0 \\
		\vdots & \vdots & \ddots & \vdots & \vdots \\
		0 & 0 & \cdots & \cos \theta_d & -\sin \theta_d \\
		0 & 0 & \cdots & \sin \theta_d & \cos \theta_d
	\end{pmatrix}
\end{equation}
and then measuring the \( \op{x}_2^{(i)} \) quadratures.
The resultant probability distributions are Gaussian with first-order moments
\begin{equation}
	\vec{w} = \frac{2h}{h_{\text{SQL}}}
		\begin{pmatrix} \sqrt{\kappa_1} \cos \theta_1 \\ \sqrt{\kappa_2} \cos \theta_2 \\ \vdots \\ \sqrt{\kappa_d} \cos \theta_d \end{pmatrix},
\end{equation}
and the second-order moments are given by
\begin{equation}
	\Sigma =
	(1-\eta) \identity + \eta \left( FQF + FSG + GSF + GRG - \left( GS + FQ \right) L G - G L \left( SG + QF \right) + K_{\text{Tot}} GLG \right),
\end{equation}
where
\(
	G_{ij} = \delta_{ij} \cos \theta_i,
\)
and
\(
	F_{ij} = \delta_{ij} \sin \theta_i.
\)
The \ac{CFI} of a Gaussian probability distribution can similarly be evaluated in terms of its moments as~\cite{kay_fundamentals_1998}
\begin{equation}
	\mathcal{I}(h)
	=
	2\frac{ \partial \vec{w}^T }{ \partial h } \Sigma^{-1} \frac{ \partial \vec{w} }{ \partial h }
	+
	\frac{1}{2}\trace{ \left( \frac{ \partial \Sigma }{ \partial h } \Sigma^{-1} \right)^2 },
	\label{eq:cfi_gaussian}
\end{equation}
where our definition of the covariance matrix \( \sigma \) differs from that of Ref.~\cite{kay_fundamentals_1998} by a factor of \( 2 \), leading to the extra factor in the first term of Eq.~\eqref{eq:cfi_gaussian}.
The \ac{CFI} is then given by
\[
	\mathcal{I}(h)=
	\frac{8\eta}{h_{\text{SQL}}^2}
	\sum\limits_i
	\cos \theta_i \cos \theta_j \sqrt{\kappa_i \kappa_j}
	\left( \Sigma^{-1} \right)_{ij}
	=
	\frac{8\eta}{h_{\text{SQL}}^2}
	\expect{G^2\Sigma^{-1}},
\]
where \( \Sigma^{-1} \) can be calculated with Eq.~\eqref{eq:woodbury} to be
\begin{equation}
\begin{aligned}
	\Sigma^{-1}
	&=
	Y^{-1}
	\\
	&\mkern32mu -
	Y^{-1}
	\begin{pmatrix}
		G \vec{k} &
		\left( G S + F Q \right) \vec{k}
	\end{pmatrix}
	\begin{pmatrix}
		\expect{G^2 Y^{-1}} & -\frac{1}{\eta} + \expect{G Y^{-1} \left( G S + F Q \right)} \\
		-\frac{1}{\eta} + \expect{G Y^{-1} \left( G S + F Q \right)} & -\frac{(1-\eta)}{\eta}\expect{Q Y^{-1}} - \expect{G^2 Y^{-1}}
	\end{pmatrix}^{-1}
	\begin{pmatrix}
		\vec{k}^T G \\
		\vec{k}^T \left( S G + Q F \right)
	\end{pmatrix}
	Y^{-1},
\end{aligned}
\end{equation}
where \( Y = (1-\eta)\identity + \eta \left( F^2 Q + 2 F G S + G^2 R \right) \).
Thus \(\mathcal{I} \) is given by
\begin{equation}
	\begin{aligned}
		\mathcal{I}(h)
		&=
		\frac{8\eta}{h_{\text{SQL}}^2}
		\expect{G^2 \Sigma^{-1}}
		\\
		&=
		\frac{8\eta}{h_{\text{SQL}}^2}
		\Bigg[\expect{G^2 Y^{-1}}
			-
			\begin{pmatrix}
				\expect{G^2 Y^{-1}} & \expect{G Y^{-1} \left( G S + F Q \right)}
			\end{pmatrix}
			\\
	&\mkern128mu
	\begin{pmatrix}
		\expect{G^2 Y^{-1}} & -\frac{1}{\eta} + \expect{G Y^{-1} \left( G S + F Q \right)} \\
		-\frac{1}{\eta} + \expect{G Y^{-1} \left( G S + F Q \right)} & -\frac{(1-\eta)}{\eta}\expect{Q Y^{-1}} - \expect{G^2 Y^{-1}}
	\end{pmatrix}^{-1}
\begin{pmatrix}
	\expect{G^2 Y^{-1}} \\
	\expect{G Y^{-1} \left( G S + F Q \right)}
\end{pmatrix}
		\Bigg]
\\
&=
\frac{8\eta}{h_{\text{SQL}}^2}
\frac{\expect{G^2 Y^{-1}}}{\left[ 1 - \eta\expect{G Y^{-1} (G S + F Q)} \right]^2 + \eta \expect{G^2 Y^{-1}}\left[ (1-\eta)\expect{Q Y^{-1}} + \eta \expect{G^2 Y^{-1}}\right]}.
	\end{aligned}
	\label{eq:app:squeezed_hom_general}
\end{equation}
which gives a \ac{CRB} of
\begin{equation}
	\left( \Delta h \right)^2
	\geq
	\frac{h_{\text{SQL}}^2}{8}
	\left\{
		\frac{\left[ 1-\eta\left( \expect{G^2 Y^{-1} S} + \expect{F G Y^{-1} Q} \right) \right]^2}{\eta\expect{G^2 Y^{-1}}} + (1-\eta) \expect{Q Y^{-1}} + \eta \expect{G^2 Y^{-1}}
	\right\}.
	\label{eq:app:hom_crb}
\end{equation}

\subsection{Cramér-Rao bounds for measurement along the signal quadratures}
For homodyne measurements along the signal quadrature Eq.~\eqref{eq:app:squeezed_hom_general} reduces to
\begin{equation}
\begin{aligned}
	\mathcal{I}
	&=
	\frac{8\eta}{h_{\text{SQL}}^2}
	\expect{\Sigma^{-1}}
	\\
	&=
	\frac{8\eta}{h_{\text{SQL}}^2}
	\left[\expect{T^{-1}}
		-
		\begin{pmatrix}
			\expect{T^{-1}} & \expect{T^{-1} S}
		\end{pmatrix}
	\begin{pmatrix}
		\expect{T^{-1}} & -\frac{1}{\eta} + \expect{T^{-1} S} \\
		-\frac{1}{\eta} + \expect{T^{-1} S} & -\frac{K_{\text{Tot}}}{\eta} + \expect{S T^{-1} S}
	\end{pmatrix}^{-1}
\begin{pmatrix}
	\expect{T^{-1}} \\
	\expect{T^{-1} S}
\end{pmatrix}
\right]
\\
&=
\frac{8\eta}{h_{\text{SQL}}^2}
\frac{\expect{T^{-1}}}{\left( 1 - \eta\expect{T^{-1}S} \right)^2 + \eta \expect{T^{-1}}\left( \expect{Q} - \eta \expect{S T^{-1} S} \right)},
\end{aligned}
\end{equation}
where the limit is given by \( \theta_i = 0, \forall i\), which implies \( F = 0 \), \( G = \identity \), and \( Y = T \).
With \ac{CRB}
\begin{equation}
	\left( \Delta h \right)^2 \geq
\frac{h_{\text{SQL}}^2}{8}
\left[ \frac{ ( 1 - \eta \expect{T^{-1}S} )^2}{\eta\expect{T^{-1}}} + \expect{Q} - \eta \expect{ST^{-1}S}  \right]
	\label{eq:app:signalhom_crb}
\end{equation}

In the lossless regime the bound becomes
\begin{equation}
	(\Delta h)^2 \geq
	\frac{h_{\text{SQL}}^2}{8}
	\left[
	\frac{\left( 1 - \expect{R^{-1}S} \right)^2}{\expect{R^{-1}}} + \expect{R^{-1}}
\right].
\label{eq:app:lossless_signalhom_crb}
\end{equation}

\subsection{Cramér-Rao bounds for measurement along the optimal quadrature}
\label{app:hom_optimal}
In the limit \( r_i = r, \forall i \), \( \phi_i = \phi, \forall i \), and \( \theta_i = \theta, \forall i \); Eq.~\eqref{eq:app:hom_crb} reduces to
\begin{equation}
\begin{aligned}
	\left( \Delta h \right)^2
	\geq
	\frac{h_{\text{SQL}}^2}{8\eta}
	\Bigg[&
		\frac{(1-\eta)\sec^2 \theta + \eta \left( \cosh 2r - \sinh 2r \cos 2\phi + 2\sinh 2r \sin 2\phi \tan \theta + \left( \cosh 2r + \sinh 2r \cos 2\phi \right) \tan^2 \theta \right)}{\kappa_{\text{Tot}}}
		\\
		&\mkern32mu
		+ \eta \left( K_{\text{Tot}}
		- 2\left( \sinh 2r \sin 2\phi+ \left( \cosh 2r + \sinh 2r \cos 2\phi \right) \tan \theta \right)
		\right)
	\Bigg].
\end{aligned}
\label{eq:app:hom_signal_crb}
\end{equation}
The optimal homodyne angle for this case is given by
\begin{equation}
	\theta =
	\arctan \left(
		\eta
		\frac{K_{\text{Tot}} - \sinh 2r \sin 2\phi}{ 1-\eta + \eta ( \cosh 2r + \sinh 2r \cos 2\phi )}
	\right),
\end{equation}
at which homodyne angle the \ac{CRB} becomes
\begin{equation}
	\left( \Delta h \right)^2
	\geq
	\frac{h_{\text{SQL}}^2}{8\eta}
	\frac{
		\eta^2
		+
		(1-\eta)
		\left[
		\left( 1-\eta \right) + 2\eta \cosh 2r
		-2\kappa_{\text{Tot}} \sinh 2r\sin 2\phi
		+\eta \kappa_{\text{Tot}} K_{\text{Tot}}
	\right]
}{(1-\eta) \kappa_{\text{Tot}} + \eta K_{\text{Tot}}},
\end{equation}
attaining the \ac{QCRB} given in Eq.~\eqref{eq:app:qcrb_loss_equal_squeeze}.

In the lossless limit \( \eta \to 1 \) (with squeezings unconstrained) the \ac{CRB} becomes
\begin{equation}
	\left( \Delta h \right)^2
	\geq
	\frac{h_{\text{SQL}}^2}{8}
	\left\{
		\frac{\left[ 1-\left( \expect{G^2 Z^{-1} S} + \expect{F G Z^{-1} Q} \right) \right]^2}{\expect{G^2 Z^{-1}}} + \expect{G^2 Z^{-1}}
	\right\},
	\label{eq:app:lossless_hom_crb}
\end{equation}
where \( Z = F^2 Q + 2 F G S + G^2 R \).
The homodyne angle
\begin{equation}
	\theta_i = \arctan
	\left( \frac{K_{\text{Tot}} - \sinh 2r_i \sin 2\phi_i}{\cosh 2r_i + \sinh 2r_i \cos 2\phi_i} \right),
	\label{eq:app:hom_angle_lossless}
\end{equation}
then attains the \ac{QCRB} of
\begin{equation}
	\left( \Delta h \right)^2
	\geq
	\frac{h_{\text{SQL}}^2}{8}
	\frac{1}{K_{\text{Tot}}}.
\end{equation}

\section{Multi-carrier optimum sensitivity}
The Cramér-Rao bounds Eqs.~\eqref{eq:app:qcrb_loss_squeeze},~\eqref{eq:app:hom_crb}, and~\eqref{eq:app:signalhom_crb} all have form
\begin{equation}
	\frac{8}{h_{\text{SQL}}^2}\left( \Delta h \right)^2 \geq B(\kappa_1,\cdots,\kappa_d) = \frac{\Big( 1 - \sum\limits_i c_1^{(i)} \kappa_i \Big)^2}{\sum\limits_i c_2^{(i)} \kappa_i} + \sum\limits_i c_3^{(i)} \kappa_i,
	\label{eq:app:multi_abstract}
\end{equation}
with \( c_2^{(i)} \geq 0 \) and \( c_3^{(i)} \geq 0 \) with equality only holding only (but not necessarily) if \( \eta = 1 \).

The optimum sensitivity is achieved by minimising \( B \) which---unlike typical interferometry cases---generally does not find maximum at \( \kappa_i \to \infty \) as \( c_3^{(i)} > 0 \) means \( B \) diverges when any \( \kappa_i \to \infty \) and similarly at \( \kappa_i = 0, \forall i \) \( B \) diverges.
The derivatives and Hessian of Eq.~\eqref{eq:app:multi_abstract} are
\begin{gather}
	\frac{\partial B}{\partial \kappa_j} =  -c_2^{(j)} \left( \frac{ 1 - \sum\limits_i c_1^{(i)} \kappa_i }{\sum\limits_i c_2^{(i)} \kappa_i} \right)^2 - 2c_1^{(j)} \left( \frac{ 1 - \sum\limits_i c_1^{(i)} \kappa_i }{\sum\limits_i c_2^{(i)} \kappa_i} \right) + c_3^{(j)},
	\label{eq:app:B_grad} \\
	\frac{\partial^2 B}{\partial \kappa_j \partial \kappa_k} =
	\frac{2}{\sum\limits_i c_2^{(i)} \kappa_i} 
	\left[ c_1^{(j)} + c_2^{(j)} \left( \frac{ 1 - \sum\limits_i c_1^{(i)} \kappa_i }{\sum\limits_i c_2^{(i)} \kappa_i} \right) \right]
	\left[ c_1^{(k)} + c_2^{(k)} \left( \frac{ 1 - \sum\limits_i c_1^{(i)} \kappa_i }{\sum\limits_i c_2^{(i)} \kappa_i} \right) \right],
	\label{eq:app:B_hessian}
\end{gather}
which being positive semi-definite for \( \sum\limits_i c_2^{(i)} \kappa_i > 0 \) identifies \( B \) as convex.

Solving for roots of Eq.~\eqref{eq:app:B_grad} we find 
\begin{equation}
	\frac{ 1 - \sum\limits_i c_1^{(i)} \kappa_i }{ \sum\limits_i c_2^{(i)} \kappa_i } 
	=
	- \frac{c_1^{(j)}}{c_2^{(j)}} \pm \sqrt{\left(\frac{c_1^{(j)}}{c_2^{(j)}}\right)^2 + \frac{c_3^{(j)}}{c_2^{(j)}} },
	\label{eq:app:B_grad_soln}
\end{equation}
where the \( + \) and \( - \) solutions are valid for \( \sum\limits_i c_1^{(i)} \kappa_i < 1 \) and \( \sum\limits_i c_1^{(i)} \kappa_i > 1 \) respectively; and are always positive and negative respectively.
This indicates that a solution to \( \partial_{\kappa_1} B = \cdots = \partial_{\kappa_d} B = 0 \) cannot be found unless
\begin{equation}
	- \frac{c_1^{(j)}}{c_2^{(j)}} + \sqrt{\left(\frac{c_1^{(j)}}{c_2^{(j)}}\right)^2 + \frac{c_3^{(j)}}{c_2^{(j)}} }
	=
	- \frac{c_1^{(k)}}{c_2^{(k)}} + \sqrt{\left(\frac{c_1^{(k)}}{c_2^{(k)}}\right)^2 + \frac{c_3^{(k)}}{c_2^{(k)}} },
	\label{eq:app:multiroot_plus}
\end{equation}
for all \( j \) and \( k \); or
\begin{equation}
	- \frac{c_1^{(j)}}{c_2^{(j)}} - \sqrt{\left(\frac{c_1^{(j)}}{c_2^{(j)}}\right)^2 + \frac{c_3^{(j)}}{c_2^{(j)}} }
	=
	- \frac{c_1^{(k)}}{c_2^{(k)}} - \sqrt{\left(\frac{c_1^{(k)}}{c_2^{(k)}}\right)^2 + \frac{c_3^{(k)}}{c_2^{(k)}} },
	\label{eq:app:multiroot_minus}
\end{equation}
for all \( j \) and \( k \).

Considering first the case when neither Eq.~\eqref{eq:app:multiroot_plus} nor Eq.~\eqref{eq:app:multiroot_minus} hold for any \( j \) or \( k \).
As \( \kappa_i = 0, \forall i \) and \( \exists i | \kappa_i = \infty \) lead \( B \) to diverge, \( B \) must be extremised for \( \kappa_i \in [0,\infty) \) with \( \sum\limits_i \kappa_i \in (0,\infty) \).
If \( \partial_{\kappa_i} B = 0\) and \( \partial_{\kappa_j} B = 0 \) cannot be simultaneously satisfied then we find instead a set of possible minima \( \{ \kappa_i | \kappa_j = 0, \forall j\neq k\} \) where \( \kappa_k \) is such that \( \partial_{\kappa_k} B = 0 \).
These solutions are exactly the single-carrier configurations for the interferometer, which have optimal precision at
\begin{equation}
	\kappa_i = \pm \frac{1}{\sqrt{(c_1^{(i)})^2 + c_2^{(i)} c_3^{(i)}}},
\end{equation}
where the \( - \) solution can be discarded as unphysical.
The overall optimum sensitivity is thus
\begin{equation}
	\min\limits_i \left\{ -2\frac{c_1^{(i)}}{c_2^{(i)}} + 2\sqrt{\left(\frac{c_1^{(i)}}{c_2^{(i)}}\right)^2 + \frac{c_3^{(i)}}{c_2^{(i)}} } \right\},
\end{equation}
which is obtained using the carrier \( l \) where
\begin{equation}
	l = \argmin\limits_i \left\{ -2\frac{c_1^{(i)}}{c_2^{(i)}} + 2\sqrt{\left(\frac{c_1^{(i)}}{c_2^{(i)}}\right)^2 + \frac{c_3^{(i)}}{c_2^{(i)}} } \right\}.
\end{equation}

When the solutions to \( \partial_{\kappa_i} B = 0 \) and \( \partial_{\kappa_j} B = 0 \) are compatible with one another then a family of potential solutions exist satisfying
\begin{equation}
	\frac{1 - \sum\limits_{i \in \mathcal{A}_j^{\pm}} c_1^{(i)} \kappa_i}{\sum\limits_{i \in \mathcal{A}_j^{\pm}} c_2^{(i)} \kappa_i}
	=
	-\frac{c_1^{(i)}}{c_2^{(i)}} \pm \sqrt{\left(\frac{c_1^{(i)}}{c_2^{(i)}}\right)^2 + \frac{c_3^{(i)}}{c_2^{(i)}} },
\end{equation}
which reduces to
\begin{equation}
	\sum\limits_{i \in \mathcal{A_j^{\pm}}} \sqrt{(c_1^{(j)})^2 + c_2^{(j)}c_3^{(j)}} \kappa_j = \pm 1,
	\label{eq:app:degenerate_set}
\end{equation}
where
\begin{equation}
	\mathcal{A}_j^{\pm} =
	\left\{ i \middle| 
	- \frac{c_1^{(j)}}{c_2^{(j)}} \pm \sqrt{\left(\frac{c_1^{(j)}}{c_2^{(j)}}\right)^2 + \frac{c_3^{(j)}}{c_2^{(j)}} }
	=
	- \frac{c_1^{(i)}}{c_2^{(i)}} \pm \sqrt{\left(\frac{c_1^{(i)}}{c_2^{(i)}}\right)^2 + \frac{c_3^{(i)}}{c_2^{(i)}} }
	\right\}.
\end{equation}
The \( - \) solutions again require some \( \kappa_i < 0 \) which is unphysical and so we can consider only the positive solutions.
When solutions \( \{ \kappa_i \} \) satisfy Eq.~\eqref{eq:app:degenerate_set} the precision attained is
\begin{equation}
	-2\frac{c_1^{(i)}}{c_2^{(i)}} + 2\sqrt{\left(\frac{c_1^{(i)}}{c_2^{(i)}}\right)^2 + \frac{c_3^{(i)}}{c_2^{(i)}} },
\end{equation}
which is identical for all \( i \in \mathcal{A}_j^+ \).
For a set of configurations of which the minimum and maximum \( \kappa_{\text{Tot}} \) cases are single-carrier configurations.

\section{Applications to LIGO}
\footnotetext[1]{Values used are \( I_{\text{Tot}} = \SI{840}{\kW} \), \( L = \SI{4}{\km} \), \( M = \SI{40}{\kg} \), \( \gamma_i \approx \gamma = \SI[product-units=single]{2\pi x 500}{\per\s} \), \( \omega_i \approx \omega = \SI[product-units=single]{2\pi x 2.82e14}{\per\s} \)}
In the absence of external squeezing the fundamental limit (Eq.~(20) in the main text) is minimised by \( \kappa_{\text{Tot}} = \left[ (1-\eta)\eta \right]^{-\frac{1}{2}} \) while optimising the precision of measurement along the signal quadrature (Eq.~(21) in the main text) requires \( \kappa_{\text{Tot}} = \eta^{-\frac{1}{2}} \).
The latter limit is already achieved by the current LIGO interferometer around \( \Omega = \SI[product-units=single]{2\pi x 90}{\per\s} \)~\cite{note1}, while the introduction of frequency-dependent homodyne detection would increase the required intensity by \( 1/\sqrt{1-\eta} \) which for \( \eta \leq 0.99 \) is no more than an order of magnitude increase in the circulating power.

\begin{figure}[htb]
	\includegraphics{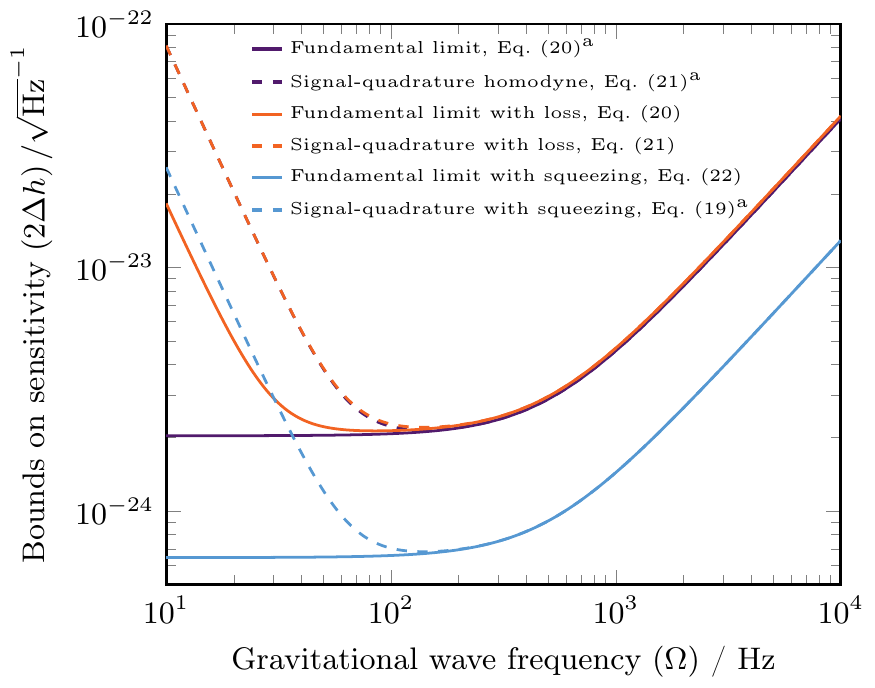}
	\footnotetext[1]{Plots are for the given equation in the \( \eta = 1 \) limit}
	\caption{
		Plots of precision attainable unsqueezed and lossless, squeezed and lossless, and unsqueezed and lossy interferometers.
		Values based on LIGO setup~\cite{danilishin_quantum_2012,note1} in the \( \kappa_{\text{Tot}} \approx g I_{\text{Tot}} \) regime, detector loss of \num{0.05} (\( \eta = 0.95 \)) and an equal squeezing amplitude \( e^{-2r} = 0.1 \) in each mode is used.
		Bounds on \( 2\Delta h \) are plotted to give equivalent values to the spectral noise density.
		Equation numbers refer to the main text.
	}
	\label{fig:plots}
\end{figure}

For the tuned gravitational-wave detector these sensitivity plots are given for identical squeezing in Fig.~\ref{fig:plots}, the case of zero external squeezing, zero loss, and zero loss and zero squeezing simultaneously with the optimal squeezing angle.
Values used are \( I_{\text{Tot}} = \SI{840}{\kW} \), \( L = \SI{4}{\km} \), \( M = \SI{40}{\kg} \), \( \gamma_i \approx \gamma = \SI[product-units=single]{2\pi x 500}{\per\s} \), \( \omega_i \approx \omega = \SI[product-units=single]{2\pi x 2.82e14}{\per\s} \)

\bibliography{tuned_multimode}